\documentclass{cargese}\usepackage{graphicx}
\let\footnote\savefootnote
\let\footnotetext\savefootnotetext 
\let\footnoterule\savefootnoterule 
\normallatexbib

\newcommand {\eqref} [1] {(\ref {#1})}
\newcommand {\beq} {\begin{equation}} 
\newcommand {\eeq} {\end{equation}}
 \newcommand {\ber}{\begin{eqnarray*}}
 \newcommand {\eer} {\end{eqnarray*}}
\newcommand {\bea}{\begin{eqnarray}}
 \newcommand {\eea} {\end{eqnarray}}

\newcommand{\ket}[1]{|#1 \rangle}

\def\cZ{{\cal Z}}
 
\def\cN{{\cal N}}

\def\tr{{\rm Tr}}

\def\vphi{\varphi}

\newcommand{\be}{\begin{equation}}
\newcommand{\ee}{\end{equation}}
\newcommand{\eq}[1]{(\ref{#1})}

\begin{document}
\articletitle
{The complete one--loop spin chain \\
for ${\cal N}=2$ Super Yang--Mills}
\chaptitlerunninghead
{Complete spin chain for ${\cal N}=2$ SYM} 
\author{Paolo Di Vecchia}
\affil{NORDITA,
Blegdamsvej 17, DK-2100 Copenhagen \O, Denmark \\ }    
\email{divecchi@alf.nbi.dk}
\author{Alessandro Tanzini}
\affil{SISSA, via Beirut 2/4, 34014 -- Trieste, Italy}
\email{tanzini@sissa.it}
\begin{abstract}
We show
that the complete planar one--loop mixing matrix of 
the ${\cal N}=2$ Super Yang--Mills theory can be obtained
from a reduction of that of the ${\cal N}=4$ theory.
For composite operators of scalar fields, this yields
an anisotropic XXZ spin chain, whose spectrum of excitations displays
a mass gap. 
\end{abstract}

\section{Introduction and discussion}

The study of the AdS/CFT correspondence in some particular subsectors
characterised by large quantum numbers, 
started with~\cite{Berenstein:2002jq}, provided a new and very intuitive way 
to look at the duality between gauge and string theory. 
The basic idea is that in presence of large quantum numbers one can
give a semiclassical description of the string states.
This semiclassical description is reproduced on the gauge theory side
by the coherent states of a quantum mechanical model, 
which can be extracted from
the study of a particular subsector of the dual gauge theory.

More precisely,
the original idea of Berenstein et al.~\cite{Berenstein:2002jq} 
was to regard  
some gauge--invariant operators of ${\cal N}=4$ Super Yang--Mills (SYM) theory 
as a discretized version of the physical type IIB string on the plane--wave 
background.
The BMN operators are single trace operators containing a large number of one 
of the complex scalar fields of ${\cal N}=4$ SYM, with the insertion of 
few \textit{impurities}
given by the other fields of the ${\cal N}=4$ supermultiplet, 
each of them corresponding to a different excitation of the string. 
The mass of the string corresponds to the anomalous dimension of the 
gauge theory operator
(see \cite{pp-rev} for a review).

It was soon realised \cite{Gubser:2002tv} that the study of type IIB string 
on the plane--wave background corresponds to a semiclassical
expansion around a particular solution of the $AdS_5\times S^5$ 
sigma model, describing
a massless point--like string spinning along the equator of $S^5$.
Then also more general solutions describing 
extended strings spinning in $AdS_5$ and/or in $S^5$ were analysed
\footnote{For an up--to--date review, see the Tseytlin's
lectures at this school \cite{Tseytlin:2004xa}.}.
The corresponding gauge theory operators bring in this case a large
number of impurities. This makes very difficult to compute their
anomalous dimensions, which are given by the eigenvalues of a 
large mixing matrix. A very interesting observation 
was made in \cite{Minahan:2002ve}, where the one--loop mixing matrix
for the operators containing scalar impurities
was identified with the Hamiltonian of an integrable 
spin--chain. 
This allows one
to compute the anomalous dimensions of the ``long''
gauge theory operators
by using the Bethe ansatz.
The integrability property has then been extended
to the full ${\cal N}=4$ dilatation operator
and studied also to higher orders in the 't Hooft coupling
(see \cite{Beisert:2004ry} for a review).

Besides giving evident computational advantages, 
the relationship with integrable systems 
brings also new insights on the gauge/string duality.
In fact the infrared dynamics of the spin--chain, for 
coherent state excitations with
wavelenghts much longer than the distance between two
nearest--neighbors sites, is described by a 
sigma model which can be mapped to that of a string
spinning in $S^5$ \cite{Kruczenski:2003gt}.
In this way, not only the mass of the string is identified
with the anomalous dimensions, but also its shape can be identified 
with the mean value of the spin on the coherent state
built with the gauge theory operators.
Moreover, these states probe regions of the string spectrum
far away from the states protected by supersymmetry.
This suggest a new possible path to extract informations
about the string dual of gauge theories,
which can be studied also in cases where some (or all)
the supersymmetries are broken and the conformal invariance is lost.

Here we discuss the one--loop renormalisation
of composite operators in ${\cal N}=2$ SYM
theory, and we observe that the {\it complete} mixing matrix
can be simply obtained from
a reduction of that of the maximally supersymmetric ${\cal N}=4$
SYM theory. Then, by focusing on the subsector of composite operators of
scalar fields, we recover the identification of their
mixing matrix 
with the Hamiltonian of an {\it anisotropic} XXZ spin chain 
\cite{DiVecchia:2004jw}.

Anisotropic (XXZ) spin--chains have been found recently also in the
study of some subsectors of ${\cal N}=1$ SYM and in pure Yang--Mills
in the light--cone gauge \cite{Belitsky:2004sc}.
Also very recently it appeared an analysis of the two--loops
dilatation operator in QCD and ${\cal N}=1$ SYM \cite{Belitsky:2004sf}.
It would be interesting to see whether the direct relationship 
between the ${\cal N}=2$ and the 
${\cal N}=4$ mixing matrix found here still persist at higher orders.
In fact the ground state of the ${\cal N}=2$ 
XXZ spin chain is protected to all orders of perturbation theory
\cite{Maggiore:2001}, and one can expect that the integrability 
property could be maintained at higher orders at least in some subsector. 

Finally, we observe that the direct relationship between
the dilatation operator of the ${\cal N}=2$ and the ${\cal N}=4$  
theory that we presented here could be a useful tool to 
explore a possible dual string theory description of 
composite operators of the ${\cal N}=2$ SYM.

\section{Operator mixing in ${\cal N}=2$ SYM}

We start by writing the Lagrangian of ${\cal{N}}=2$ Super Yang-Mills in
Weyl notations 
\begin{eqnarray}
L_{E} &=& \frac{2}{g^2} \tr\Big(\frac{1}{4}F_{\mu \nu}F_{\mu \nu} + (D_{\mu} 
\phi )^{\dagger} D_{\mu} \phi
+ \psi  \sigma^{\mu}  D^{\mu} \bar\psi 
+ \lambda \sigma^{\mu} D^{\mu}\bar\lambda
\nonumber\\
&& - i\sqrt{2}\left(
\psi [\bar\phi,\lambda] + \bar\psi  [\phi, \bar\lambda]
\right) 
+ \frac{1}{2}[\bar\phi,\phi]^2 \Big) \ ,
\label{lnee}
\end{eqnarray}
in terms of the euclidean $\sigma$-matrices 
$\sigma^{\mu}=({\bf 1},i\tau^i)$, $\tau^i$ being the Pauli matrices.
The field $\phi$ is the complex scalar field 
of the $\cN=2$ Super
Yang-Mills, the two Weyl spinors $\lambda$ and $\psi$ are the
fermionic superpartners and the covariant derivative reads 
$D_{\mu} \phi = \partial_{\mu} \phi - i [A_{\mu}, \phi]$.

We are interested in studying the planar, one--loop renormalization 
of composite operators of the elementary fields
appearing in the Lagrangian (\ref{lnee}).
We observe that the mixing matrix in this approximation
can be directly obtained from a reduction of 
that of ${\cal N}=4$ SYM, which can be found in 
\cite{Beisert:2004ry}. 
In fact, the ${\cal N}=4$ theory can be seen as 
an ${\cal N}=2$ SYM coupled with an hypermultiplet
in the adjoint representation.
The Feynman diagrams contributing
to the mixing of operators containing only 
fields of the ${\cal N}=2$ vector multiplet
are the same in the two theories, except for
the self--energy.  
However, we will show that the wave function 
renormalisation for the fields in the Lagrangian (\ref{lnee})
is exactly the same as in the ${\cal N}=4$ theory.
Thus we can read the mixing matrix of the ${\cal N}=2$ SYM
simply by restricting the indices of that 
of the ${\cal N}=4$ SYM to the ${\cal N}=2$ vector multiplet.
Similar arguments based on the inspection of the Feynman diagrams
were used to connect the one--loop light--cone mixing matrices 
of Yang--Mills theories with $0\le {\cal N} \le 4$ supersymmetry
\cite{Belitsky:2004yg,Belitsky:2004sc}
and more recently to obtain the complete one--loop mixing
of QCD composite operators in a covariant formalism \cite{Beisert:2004fv}.
 
\begin{figure}[ht]
\begin{center}
{\scalebox{1}{\includegraphics{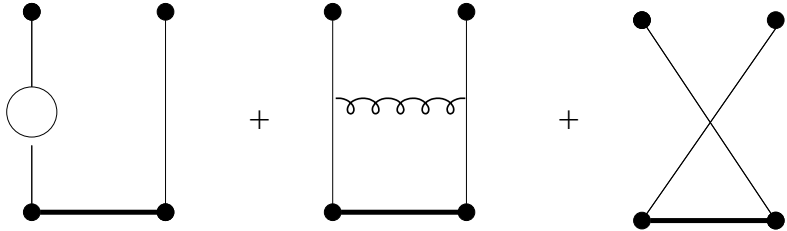}}}
\caption{Feynman diagrams contributing at one--loop. The thick horizontal
line joins the fields belonging to the composite operator. 
\label{f1}}
\end{center}
\end{figure}

Let us focus for the moment on the mixing of 
operators containing only the two real scalar fields
of the ${\cal N}=2$ vector multiplet, related to the complex field as
$\phi= 1/\sqrt{2}(\varphi_1 + i \varphi_2)$
\footnote{This mixing has been studied in \cite{DiVecchia:2004jw}.
Here we use the same conventions: the generators of the gauge group 
are normalized as $\tr (T^a T^b ) = \frac{1}{2} \delta^{ab}$
and the relations between the bare and renormalized quantities are
$g = Z_g g_r $ and $\varphi  = Z_{\varphi}^{1/2} \varphi_r$.}
\be
{\cal O}=\tr\left(\varphi_{i_{1}}\ldots \varphi_{i_{l}}
\varphi_{i_{l+1}}\ldots\varphi_{i_L}\right) \ .
\label{O-real}
\ee
We then study the correlator
\be
\langle \varphi^{i_L}\ldots\varphi^{i_{l+1}}\varphi^{i_l}\ldots\varphi^{i_1}
{\cal O}\rangle= \cZ_\varphi^{L/2}\cZ_{\cal O} 
\langle \varphi^{i_L}_r\ldots\varphi^{i_{l+1}}_r\varphi^{i_l}_r
\ldots\varphi^{i_1}_r
{\cal O}_r\rangle \ ,
\label{corr-real}
\ee
where $\cZ_{\varphi}$ is the usual wave--function renormalization
needed to make finite the two--point function 
$\langle\bar\varphi_r (x)\varphi_r (y)\rangle$ and 
$\cZ_{{\cal O}}$ is the renormalization factor for the composite
operator.
The operators (\ref{O-real}) mix among themselves at the quantum
level, and ${\cal Z}_{\cal O}$ is a matrix carrying the indices of the 
real fields. Actually the operators (\ref{O-real}) mix at one--loop
also with operators containing derivatives of the scalar fields,
but the mixing matrix is triangular \cite{DiVecchia:2004jw}.
Then, for the computation of the anomalous dimensions 
we can neglect this mixing and study only the correlators (\ref{corr-real}).
By using the large $N$ approximation, we focus  on the 
nearest--neighbors interaction
\begin{equation}
\langle\ldots\vphi_{i_{l+1}}(x)\varphi_{i_l}(y)
\ldots\tr\Big(\ldots\vphi_{j_l}\varphi_{j_{l+1}}\ldots\Big)(z)\rangle
\label{nn-real}
\end{equation}
The corresponding one--loop diagrams are displayed in 
Fig.(\ref{f1}). 

Concerning the first,
it turns out that in ${\cal{N}}=2$ Super Yang-Mills all the self-energy 
diagrams cancel. This means that in the convention we choose for the
lagrangian \eq{lnee} the only renormalization for the fields is that associated
to the gauge coupling, {\it i.e.} $\cZ_{\varphi}^{1/2}\equiv\cZ_g$.
On the other hand from the knowledge of the 
$\beta$-function
\begin{equation}
\beta(g) \equiv \mu\frac{\partial}{\partial\mu}g_r = 
-g \mu\frac{\partial}{\partial\mu} \log \cZ_g =
-\frac{g^3}{16\pi^2}2N
\label{beta2}
\end{equation}
one can derive the expression for $\cZ_g$
\begin{equation}
\cZ_g  = 1 - \frac{g^2 N}{16\pi^2} \frac{\mu^{-2\epsilon}}{\epsilon} \ .
\label{zg2}
\end{equation}
Indeed \eq{beta2} follows from \eq{zg2} after taking the $\epsilon\to 0$ limit.
Then we get 
\be
\cZ_{\varphi} = 1 - \frac{g^2N}{4\pi^2}\frac{\mu^{-2\epsilon}}{2\epsilon}
\label{zphi-n2}
\ee
giving $\gamma_\varphi=1/2\mu \partial/\partial_\mu \cZ_{\varphi}=g^2 N/8\pi^2$
exactly as for the ${\cal N}=4$ theory in the Feynman gauge 
\cite{Erickson:2000af}.  
The $\cZ_\varphi^{L/2}$ factor in (\ref{corr-real}) takes care  
of half of the contribution associated to the wave function renormalisation.
We have thus to multiply the composite operator by a factor
\begin{equation}
 \cZ^{(g)\ldots j_l j_{l+1}\ldots}_{\ \ \ldots i_l i_{l+1}\ldots}= {\bf 1}  
- \frac{g^2N}{8\pi^2}\frac{\mu^{-2\epsilon}}{2\epsilon}\
\delta_{i_l}^{j_l}\delta_{i_{l+1}}^{j_{l+1}} \ \ . 
\label{Z-g}
\end{equation}
for each nearest--neighbor.
    
Coming now to the other two diagrams of Fig.(\ref{f1}), 
the one--loop correction associated to the gluon exchange is\footnote{For details on the computations, we refer to \cite{DiVecchia:2004jw}.}
\begin{equation}
\cZ^{(gluon)\ldots j_l j_{l+1}\ldots}_{\ \ \ldots i_l i_{l+1}\ldots}= {\bf 1}  
+ \frac{g^2N}{16\pi^2}\frac{\mu^{-2\epsilon}}{2\epsilon}\
\delta_{i_l}^{j_l}\delta_{i_{l+1}}^{j_{l+1}} \ \ ,
\label{Z-gluon}
\end{equation}
and that associated to the four--scalar interaction 
\begin{equation}
\cZ^{(four \ sc.)\ldots j_l j_{l+1}\ldots}_{\ \ 
\ldots i_l i_{l+1}\ldots}= {\bf 1}  
+ \frac{g^2N}{16\pi^2}\frac{\mu^{-2\epsilon}}{2\epsilon}\
\Big( 2 \delta_{i_l}^{j_{l+1}}\delta^{j_l}_{i_{l+1}}
- \delta_{i_l}^{j_{l}}\delta_{i_{l+1}}^{j_{l+1}}
- \delta_{i_l i_{l+1}}\delta^{j_l j_{l+1}} \Big) \ .
\label{Z-four}
\end{equation}
As anticipated, 
the contributions coming from the gluon exchange and
from the four-scalar interaction are the same as in the ${\cal{N}}=4$
case~\cite{Minahan:2002ve} except that now the indices $i_l , i_{l+1} , j_l ,
j_{l+1}$ run only over two values and not six because ${\cal{N}}=2$ Super
Yang-Mills has only two real scalars. 
Adding the three contributions in (\ref{Z-g}), (\ref{Z-gluon})
and (\ref{Z-four}) we get
\begin{equation}
{\cal{Z}}^{\ldots j_l j_{l+1}\ldots}_{\ \ \ldots i_l i_{l+1}\ldots}= {\bf 1} -
\frac{g^2 N}{16 \pi^2} \frac{\mu^{-2\epsilon}}{2\epsilon} \left( 
\delta_{i_l i_{l+1}}\delta^{j_l j_{l+1}}  + 2 
\delta_{i_l}^{j_l}\delta_{i_{l+1}}^{j_{l+1}} - 
 2 \delta_{i_l}^{j_{l+1}}\delta^{j_l}_{i_{l+1}}\right) \ .
\label{anodim}
\end{equation}
In conclusion, we see that matrix of anomalous dimensions for the  
operators (\ref{O-real}) can be directly obtained 
from that of
$\cN=4$ theory \cite{Minahan:2002ve}
by restricting its indices to that of the two real scalar fields
of the  ${\cal N}=2$
vector multiplet,  $i=1,2$~\footnote{Another difference is that 
 the 't Hooft coupling that appears in 
 \eq{anodim} is the renormalised running coupling
 $\lambda_r=g_r^2N$. However, the substitution $\lambda\to\lambda_r$
 induces only higher order corrections. With this remark in mind,
 we will write our results in terms of the bare coupling
 $\lambda$ to simplify the notation.}. 
The same applies to the mixing matrix of operators 
containing gluons and fermions.
In fact, if we consider in ${\cal N}=4$ 
the matrix elements with indices running only
on the fields of the 
${\cal N}=2$ vector multiplet, by definition
the contribution of the extra particles in the 
hypermultiplet can only appear in loops, since
they cannot appear as external states. In particular
at the one--loop level they
only appear in the self--energy diagrams.
We remark that 
the wave function renormalisation for the gluon
and the fermion fields
is the same of that of the scalar field (\ref{zphi-n2}),
as one expects from supersymmetry\footnote{In general
the supersymmetry can be broken when using the dimensional
regularisation scheme, but in this case 
one can check with an explicit computation
that the self--energies are all the same (in particular they are
all vanishing for ${\cal N}=2$).}.
The remaining Feynman diagrams are the same in ${\cal N}=4$
and ${\cal N}=2$ theories.
Then we conclude that the complete one--loop mixing matrix 
of composite operators in ${\cal N}=2$ SYM can be directly read
from that of the ${\cal N}=4$ theory. 

\section{Scalar operators and the XXZ spin chain}

Let us now come to the relation with the spin--chain, focusing
on the sector of operators (\ref{O-real}).
Quite naturally the two scalar fields of the ${\cal N}=2$ SYM can be 
interpreted as different
orientations of a spin and then the whole gauge invariant operator
formed just by scalars can be seen as a spin chain. The cyclicity of
the trace makes the chain closed 
and implies that the physical states of the chain
corresponding to the gauge theory operators have zero total momentum.
Before to write down the spin chain Hamiltonian we
observe that the operators containing only products
of the complex scalar field 
$\phi$ 
have vanishing anomalous dimensions.
In fact these operators, when represented
in terms of the real fields $\varphi_i$, are symmetric
and traceless in the real indices $i=1,2$, and this ensures the vanishing
of their one--loop anomalous dimensions computed from \eq{anodim}.
This suggest to take them as the ground state of the spin chain
and to identify the two orientations of a spin with the following
$2$-vectors
\beq
\bar\phi \to \ket{+} \equiv
\left(
\begin{array}{c}
1\\ 0
\end{array} 
\right) ~~,~~~~
\phi \to \ket{-} \equiv
\left(
\begin{array}{c}
0\\ 1
\end{array} 
\right)~.
\label{spin}
\eeq
In this basis, the matrix of anomalous dimensions 
$\gamma_{\cal O} \equiv \cZ_{\cal O}^{-1} \mu 
\frac{\partial}{\partial\mu}\cZ_{\cal O}$, with $\cZ_{\cal O}$
given by (\ref{anodim})
reads \cite{DiVecchia:2004jw}
\be
\gamma_{\cal O} = \frac{\lambda}{16\pi^2} H_{\rm XXZ} \ ,
\label{gaano34}
\ee 
where
\be
H_{\rm XXZ}= -\frac{1}{2} \sum_{l=1}^{L} \left[ 
(\sigma^x )_{l} ( \sigma^x )_{l+1} + 
(\sigma^y )_{l} ( \sigma^y )_{l+1} +
\Delta \left( (\sigma^z)_{l} ( \sigma^z )_{l+1} - {\bf 1}_l {\bf 1}_{l+1} \right)
\right]
\label{xxz}
\ee
is the Hamiltonian of an XXZ spin chain.
For the ${\cal N}=2$ theory the value of the anisotropy parameter
is $\Delta=3$.
As anticipated, the ground state
of the spin chain corresponds to the protected operator 
${\cal O}_{vac}\equiv\tr(\phi^L)$.
The excited states are associated to spin flips along the chain,
which in the field theory language correspond to the 
insertion of ``impurities''$\bar\phi$ in the operator ${\cal O}_{vac}$. 
In \cite{DiVecchia:2004jw} we studied the spectrum associated to these excitations. The energy associated to one impurity turns out to be
\be
E_n =  \frac{\lambda}{8\pi^2} \left[ (\Delta -1) + 
\frac{2\pi^2}{L^2}n^2 \right]
\label{spectrum}
\ee
We thus see that the presence of 
a non--trivial anisotropy parameter $\Delta>1$ implies the presence of a {\it mass gap}
of the order of the 't Hooft coupling $\lambda$ in the spectrum.

\begin{acknowledgments}
It is a pleasure to thank 
the organizers of the Carg\`ese 2004 ASI. 
Part of the A.T. work was supported by the
European Union under RTN contract HPRN-CT-2000-00131. 

\end{acknowledgments}


\begin{chapthebibliography}{99}

\bibitem{Berenstein:2002jq}
D.~Berenstein, J.~M. Maldacena, and H.~Nastase,
\newblock JHEP {\bf 04}, 013 (2002), hep-th/0202021.

\bibitem{pp-rev}
A.~Pankiewicz,
\newblock Fortsch. Phys. {\bf 51}, 1139 (2003), hep-th/0307027;
J.~C. Plefka,
\newblock (2003), hep-th/0307101;
J.~M. Maldacena,
\newblock (2003), hep-th/0309246;
D.~Sadri and M.~M. Sheikh-Jabbari,
\newblock (2003), hep-th/0310119;
R.~Russo and A.~Tanzini,
\newblock Class. Quant. Grav. {\bf 21}, S1265 (2004), hep-th/0401155.

\bibitem{Gubser:2002tv}
S.~S. Gubser, I.~R. Klebanov, and A.~M. Polyakov,
\newblock Nucl. Phys. {\bf B636}, 99 (2002), hep-th/0204051.

\bibitem{Tseytlin:2004xa}
A.~A.~Tseytlin,
hep-th/0409296.

\bibitem{Minahan:2002ve}
J.~A. Minahan and K.~Zarembo,
\newblock JHEP {\bf 03}, 013 (2003), hep-th/0212208.

\bibitem{Beisert:2004ry}
N.~Beisert,
Phys.\ Rept.\  {\bf 405} (2005) 1, hep-th/0407277.

\bibitem{Kruczenski:2003gt}
M.~Kruczenski,
Phys.\ Rev.\ Lett.\  {\bf 93} (2004) 161602, hep-th/0311203;
M.~Kruczenski, A.~V.~Ryzhov and A.~A.~Tseytlin,
Nucl.\ Phys.\ B {\bf 692} (2004) 3, hep-th/0403120.

\bibitem{Belitsky:2004sc}
A.~V.~Belitsky, S.~E.~Derkachov, G.~P.~Korchemsky and A.~N.~Manashov,
hep-th/0409120.

\bibitem{Beisert:2004fv}
N.~Beisert, G.~Ferretti, R.~Heise and K.~Zarembo,
hep-th/0412029.

\bibitem{Belitsky:2004sf}
A.~V.~Belitsky, G.~P.~Korchemsky and D.~Muller,
hep-th/0412054.

\bibitem{Maggiore:2001}
A.~Blasi {\em et~al.},
\newblock JHEP {\bf 05}, 039 (2000), hep-th/0004048;
V.~E.~R. Lemes, M.~S. Sarandy, S.~P. Sorella, A.~Tanzini, and O.~S. Ventura,
\newblock JHEP {\bf 01}, 016 (2001), hep-th/0011001;
V.~E.~R. Lemes {\em et~al.},
\newblock (2000), hep-th/0012197;
N.~Maggiore and A.~Tanzini,
\newblock Nucl. Phys. {\bf B613}, 34 (2001), hep-th/0105005.

\bibitem{Belitsky:2004yg}
A.~V.~Belitsky, S.~E.~Derkachov, G.~P.~Korchemsky and A.~N.~Manashov,
Phys.\ Lett.\ B {\bf 594} (2004) 385, hep-th/0403085.

\bibitem{DiVecchia:2004jw}
P.~Di Vecchia and A.~Tanzini,
to appear on J. \ Geom. \ Phys., hep-th/0405262.

\bibitem{Erickson:2000af}
J.~K.~Erickson, G.~W.~Semenoff and K.~Zarembo,
Nucl.\ Phys.\ B {\bf 582} (2000) 155, hep-th/0003055.

\end{chapthebibliography}
\end{document}